\documentclass[11pt,preprintnumbers,aps,amssymb,nofootinbib,amsmath,superscriptaddress,notitlepage,prc]
{revtex4-1}
\usepackage{epsfig,epsf}
\usepackage{bm} 
\usepackage{color} 
\usepackage{slashed}
\usepackage{relsize}	
\usepackage{soul} 
\usepackage{hyperref}
\usepackage{tensor} 
\newcommand{\beq}{\begin{equation}}
\newcommand{\beql}[1]{\begin{equation}\label{#1}}
\newcommand{\eeq}{\end{equation}}
\def\bal#1\gal{\begin{align}#1\end{align}}
\newcommand{\ball}[1]{\bal\label{#1}}
\newcommand{\eq}[1]{(\ref{#1})}

\renewcommand{\b}[1]{{\bm #1}} 

\newcommand{\as}{\alpha_s}

\newcommand{\aver}[1]{\left\langle #1 \right\rangle}

\begin{document}

\title{Probing topological order of QCD at Electron Ion Collider}

\author{Kirill Tuchin}

\affiliation{
Department of Physics and Astronomy, Iowa State University, Ames, Iowa, 50011, USA}

\date{\today}


\begin{abstract}
 
The idea that the nuclear matter may posses  long range topological order is supported by the theory and the lattice calculations. At high temperature this order is instrumental in producing anomalous phenomena such as the Chiral Magnetic Effect. In the cold nuclear matter it affects the gluon distribution in the nuclear wave function at low $x$. The effect of the topological order is encapsulated in the unintegrated gluon distribution functions which are proportional, at the leading order, to the square of the gluon propagator at finite topological charge density. It is argued that  the Electron Ion Collider is well suited to study the topological order of the cold nuclear matter.

\end{abstract}

\maketitle


The search for the Chiral Magnetic Effect \cite{Kharzeev:2007tn,Kharzeev:2007jp,Fukushima:2008xe} and related phenomena that bear a witness to the topological structure of the QCD vacuum, is one of the highlights of the contemporary relativistic heavy-ion collisions program \cite{Kharzeev:2015znc}. It is believed that the matter with chiral fermions is organized into the topological domains characterized by a finite topological charge density. Moreover, it has been argued by Zhitnitsky \cite{Zhitnitsky:2013hs,Zhitnitsky:2012ej} that the correlation length of such a topological domain can be significantly larger than the typical scale of the QCD vacuum fluctuations (of the order of 1~fm) and  the topological charge is a slowly varying function of spacetime. The existence of the topological order is supported by the lattice calculations \cite{Horvath:2003yj,Horvath:2005rv,Horvath:2005cv,Alexandru:2005bn,Ilgenfritz:2007xu,Ilgenfritz:2008ia,Bruckmann:2011ve,Kovalenko:2004xm}.
 It is  a necessary condition for the experimental observation of the chiral magnetic effects because otherwise, in the absence of the long distance order,  the anomalous currents generated in different domains would average to zero even in a single event \cite{Tuchin:2018rrw}. Clearly, the topological order can manifest itself only in systems whose size significantly exceeds  the size of proton. Thus, the hot nuclear matter produced in heavy-ion collisions is a natural place to search for such topological effects. The conventional wisdom is that high temperature is advantageous for the topological effects because the rate of the sphaleron transitions increases in proportion to its fourth power. On the other hand, it is plausible that these transitions work to destroy the topological order thereby making the experimental observations more challenging. It is worth noting that the chiral magnetic effect was observed in Weyl semimetals, where the role of the topological domain is played by the separation of the Weyl nodes in the momentum space which also provides the necessary topological order \cite{Li:2014bha}.  Kang and Kharzeev proposed  to look for the topological effects in quark fragmentation \cite{Kang:2010qx}, which however is not sensitive to the topological order but rather to the typical QCD fluctuations. 
 
In this letter we propose to study the topological order of cold nuclear matter using Deep Inelastic Scattering on heavy nucleus at small $x$. The small-$x$ interactions are characterized by long coherence length $\ell_c \sim \mathrm{fm}/x$ which can exceed the size $R=A^{1/3}$~fm of a heavy nucleus. Since the typical scale of the topological charge variation is $\lambda= 1/\partial\ln\theta$, the interaction is sensitive to the topological order if $\ell_c> \lambda$ or $x<\mathrm{fm}/\lambda\ll 1$. On the other hand, at very small $x$ the nuclear wave function is populated by a large number of gluon excitations that  screen off the topological order. Therefore, only the processes with moderately small $x$, viz.\ in the range between 0.1 and 0.01, can be employed to search for the novel topological effects. In terms of the Color Glass Condensate theory \cite{Iancu:2003xm,Kovchegov:2012mbw}, the color charges of  the McLerran-Venugopalan model \cite{McLerran:1993ni,McLerran:1993ka,McLerran:1994vd,Mueller:1989st} are topologically ordered over long distances $\lambda$. The topological order of the cold nuclear matter is arguably very different from that of the Quark-Gluon Plasma. Nevertheless, it is generated by the same topological terms in the QCD Lagrangian. The topological domains of the cold nuclear matter can also be the seeds of the topological order in  Quark-Gluon Plasma generated in heavy-ion collisions \cite{Lappi:2006fp}. 

A large number of observables at small $x$ are proportional to the unintegrated gluon distribution functions defined in \eq{b1} and \eq{b2}. These distributions are sensitive to the topological order.  
In the next several paragraphs we are going to derive their explicit dependence on the parameters characterizing the topology of the nucleus.

The color field excitations in the topologically non-trivial background can be described by the Lagrangian 
\ball{a1}
L= -\frac{1}{4}G_{\mu\nu}^aG_{\mu\nu}^a+\sum_f \bar \psi_f(i\slashed{\partial}+g\slashed A)\psi_f -\theta \frac{c_A}{4}G_{\mu\nu}^a\tilde G_{\mu\nu}^a\,,
\gal
where $c_A= g^2N_f/8\pi^2$ is the chiral anomaly coefficient  \cite{Adler:1969gk,Bell:1969ts} and  $\theta$ is the pseudo-scalar field Legendre conjugate to the topological charge density. The corresponding equations of motion depend only on derivative $\partial\theta$ which is going to be denoted as\footnote{Letter $b$ is reserved for the impact parameter.} 
$\beta^\mu= (\beta_0,-\b \beta)=c_A\partial^\mu \theta= c_A(\dot \theta, -\b \nabla \theta)$. In the topologically ordered system $\theta$ is slowly varying function of spacetime so that $\beta^\mu$ is a constant vector. Its time component $\beta_0$ is proportional to the axial chemical potential, which Abelian analogue is called the chiral conductivity $\sigma_\chi$  \cite{Kharzeev:2009pj,Fukushima:2008xe} by the virtue of its special role in the Chiral Magnetic Effect. The spatial components $\b \beta$ are proportional to the splitting in momentum space of the Weyl nodes in the Weyl semimetals. 
The  Feynman gluon propagator reads \cite{Carroll:1989vb,Lehnert:2004hq}:
\ball{a4}
D_{\mu\nu}^{ab}(q)= -i \frac{q^2 g_{\mu\nu}+i\epsilon_{\mu\nu\rho \sigma}\beta^\rho q^\sigma+\beta_\mu \beta_\nu}{q^4+\beta^2 q^2-(\beta\cdot q)^2+i\varepsilon}\delta^{ab}\,.
\gal
The $\beta$-parameters enter many observable quantities  by the way of the unintegrated gluon distribution functions $\phi(x,\b k_\bot)$, which are related to the cross section of the gluon dipole on heavy nucleus \cite{Mueller:1999wm,Kovchegov:2001sc,Dominguez:2011wm} as follows
\bal
\phi^{(1)}(x,\b k_\bot)&= \frac{C_F}{8\pi^4 \alpha_s}\int d^2b \int d^2r e^{-i\b k_\bot\cdot \b r}
\left[
1-e^{-A \aver{\Gamma^{N}(\b r,\b b)}}\right]\frac{\nabla^2_r\aver{\Gamma^{N}(\b r,\b b)}}{\aver{\Gamma^{N}(\b r,\b b)}}\,, \label{b1}\\
\phi^{(2)}(x,\b k_\bot)&= \frac{C_F}{8\pi^4 \alpha_s}\int d^2b \int d^2r e^{-i\b k_\bot\cdot \b r}
\nabla^2_r
\left[
1-e^{-A \aver{\Gamma^{N}(\b r,\b b)}}\right]\,, \label{b2}
\gal
where the transverse plane is perpendicular to the nucleus momentum. The transverse vectors $\b r$ and $\b b$ denote the separation of the color dipole charges and the position of the dipole with respect to the nucleus center respectively. For example, the differential cross section for the double inclusive quark jet $q(k_1)+\bar q(k_2)$ production is proportional to  $\phi^{(1)}$ \cite{Dominguez:2011wm}:
\ball{f1} 
\frac{d\sigma^{\gamma^*A\to q\bar q+X}}{d^2P_\bot d^2k_\bot dy_1 dy_2}= \delta(x_{\gamma^*}-1) \phi^{(1)}(x_g,\b k_\bot)H\,,
\gal
where $\b P_\bot=(\b k_{1\bot}-\b k_{2\bot})/2$, $\b k_\bot= \b k_{1\bot}+\b k_{2\bot}$ and  $y_1$ and $y_2$ are the rapidities of $q$ and $\bar q$. $H$ is hard partonic amplitude which depends on $P_\bot$ and photon virtuality $Q^2$. 
The average sign in \eq{b1} and \eq{b2} refers to the average over the nucleon position since the gluon dipole can scatter on any of $A$ nucleons of the nucleus. Thus, for a nucleus of constant density $\rho$ the average scattering amplitude reads
\ball{b6}
 \aver{\Gamma^{N}(\b r,\b b)}= \frac{1}{A}\int d^2 b_a T(b_a)\, \rho\,  \Gamma^{N}(\b r,\b b')\approx \frac{1}{A}\rho T(b)\int d^2 b'  \Gamma^{N}(\b r,\b b')\,.
\gal
where $\b b_a$ is the nucleon position, $\b b'=\b b-\b b_a$ is the dipole impact parameter with respect to the nucleon ($b\approx b_a\gg b'$) and $T(b)=2\sqrt{R^2-b^2}$ for a spherical nucleus.

We now proceed with the calculation of the scattering amplitude of the gluon dipole on nucleon. The lowest order contribution $\mathcal{O}(\alpha_s)$, describing one gluon exchange, vanishes when traced over the nucleon's quark color.  The first finite contribution is the two gluon exchange at the order $\mathcal{O}(\alpha_s^2)$. Let $p$, $p'$ and $q$ be four-momenta of the dipole's gluon, proton's quark and the exchanged gluon respectively. In the eikonal approximation the quark trajectories are the straight lines so that in the center-of-mass frame $p=(p^0,\b 0,p^3)$ and $p'=(p^0,\b 0,-p^3)$ before and after the collision, while the corresponding currents are $2p^\mu$ and $2p'^\mu$, apart from the color factors. The current conservation at each vertex $p\cdot q= p'\cdot q=0$  implies that $q^0=q^3=0$. Using \eq{a4} the scattering amplitude reads
\ball{c2}
\Gamma^{N}(\b r,\b b)= \frac{\alpha_s^2}{4\pi^2}
\left|
\int d^2q_\bot \frac{q_\bot^2+i\epsilon_{ij}\beta^i_\bot  q^j_\bot-\beta_\parallel^2/2}{q_\bot^4-\beta^2q_\bot^2-(\b \beta_\bot\cdot \b q_\bot)^2}
\left( e^{-i(\b b+\b r/2)\cdot \b q_\bot}-e^{-i(\b b-\b r/2)\cdot \b q_\bot}\right)\right|^2\,,
\gal
where $\beta _\parallel^2 =\beta_0^2-\beta_3^2$ so that $\beta^2= \beta_\parallel^2-\b \beta_\bot^2$. 
Substituting \eq{c2} into \eq{b6} and integrating over the impact parameter one arrives at
\ball{c4}
 \aver{\Gamma^{N}(\b r,\b b)}=\frac{\rho T(b)}{A} \alpha_{s}^2\int d^2q_\bot
\frac{(q_\bot^2-\beta_\parallel^2/2)^2+\b \beta_\bot^2\b q_\bot^2-(\b \beta_\bot \cdot \b q_\bot)^2}{[q_\bot^4-\beta^2q_\bot^2-(\b \beta_\bot\cdot \b q_\bot)^2]^2}
\left(2- e^{-i\b r\cdot \b q_\bot}-e^{i\b r\cdot \b q_\bot}\right)\,,
\gal
In the topologically trivial background $\beta=0$  the amplitude \eq{c4} can be computed explicitly  $A\aver{\Gamma^N}= r^2Q_s^2\ln(1/r)/4$, where $Q_s^2= 4\pi \as^2 \rho T(b)$ is the saturation momentum characterizing the gluon density of nucleus \cite{Gribov:1984tu,Mueller:1999wm}.

Using \eq{c4} in \eq{b1},\eq{b2}  one obtains the corresponding unintegrated gluon distribution functions. 
At large enough $k_\bot$ the multiple scattering is a small effect and one can expand the square brackets in \eq{b1} and \eq{b2} in powers of small parameter $A\aver{\Gamma^N}\ll 1$. In this approximation, all integrals can be easily performed yielding the leading order in $\alpha_s$ contribution to the unintegrated gluon distribution functions
\ball{c6}
\phi^{(1)}(x,\b k_\bot)\approx \phi^{(2)}(x,\b k_\bot)\approx \frac{A\alpha_s C_F}{\pi^2}k_\bot^2\frac{(k_\bot^2-\beta_\parallel^2/2)^2+\b\beta^2_\bot \b k_\bot^2-(\b \beta_\bot \cdot \b k_\bot)^2}{[k_\bot^4-\beta^2k_\bot^2-(\b \beta_\bot\cdot \b k_\bot)^2]^2}\,.
\gal
This formula indicates that the gluon distribution at small $x$ is sensitive to the non-perturbative topological configurations especially in the infrared. In particular, it varies with the azimuthal angle between $\b k_\bot$ and $\b \beta_\bot$ and is finite at $k_\bot\to 0$. In the cold nuclear matter the spatial gradients $\b\beta$ are expected to be larger than the temporal one $\beta_0$ implying that $\beta$ is a spacelike vector and hence the denominator of \eq{c6} is finite at any $\b k_\bot$. 

The topological order can be studied at EIC in  the kinematic region $k_\bot \ll Q_s$, $x\sim 10^{-2}-10^{-1}$. For example, the cross section \eq{f1} is expected to be dependent on the azimuthal angle in the transverse plane. It must me stressed though that the parameters $\beta_\parallel$ and $\b \beta_\bot$ characterize the topological order in a single event, but they vary randomly between different events. This means, in particular, that the azimuthal dependence displayed in \eq{c6} is washed out when averaged over many events. The azimuthal angle correlations can be captured by considering the event-by-event fluctuations of the particle spectra. These fluctuations are expected to grow as $x$ decreases toward $x\sim 0.1$ and then decrease at even smaller $x$ as the result of the gluon saturation \cite{Gribov:1984tu}. Denoting an observable quantity proportional to an unintegrated gluon distribution function as $I(\psi)$ where $\psi$ is the azimuthal angle, i.e.\ angle between $\b\beta_\bot$ and $\b k_\bot$, one can measure the dispersion $\delta I =\aver{(I(\psi)-\aver{I(\psi)})^2}$ as a function of $x$, where the averaging is over the events.

In summary, it is argued that (i) gluon distribution functions of heavy nucleus  at low $x$  are sensitive to the topological structure of QCD and (ii)  the future Electron Ion Collider \cite{Accardi:2012qut} will study the nuclear matter in the kinematic regime that is sensitive to these novel effects. The question of whether the luminocity of EIC will be sufficient to detect these effects requires a dedicated phenomenological analysis.

\acknowledgments
I  thank A. Zhitnitsky for an enlightening discussion.
This work  was supported in part by the U.S. Department of Energy under Grant No.\ DE-FG02-87ER40371.


\end{document}